\documentclass[conference,a4paper]{APSIPA2020}\IEEEoverridecommandlockouts
\usepackage{multirow}
\usepackage{graphicx}
\usepackage{amsmath, bm}
\usepackage[psamsfonts]{amssymb}
\usepackage{amsxtra}
\usepackage{threeparttable}
\usepackage{color}
\usepackage{xcolor}
\usepackage{cite}

\begin{document}

\title{VAW-GAN for Singing Voice Conversion with Non-parallel Training Data}

\author{%
\authorblockN{%
Junchen Lu \authorrefmark{1},
Kun Zhou \authorrefmark{1},
Berrak Sisman \authorrefmark{2}\authorrefmark{1} and
Haizhou Li \authorrefmark{1}\thanks{\textbf{Codes \& Singing Samples:} \url{https://kunzhou9646.github.io/singvaw-gan/}}
}
\authorblockA{%
\authorrefmark{1}
Dept. of Electrical and Computer Engineering, National University of Singapore, Singapore}
\authorblockA{%
\authorrefmark{2}
Information Systems Technology and Design (ISTD) Pillar, Singapore University of Technology and Design, Singapore}
}

\maketitle
\thispagestyle{empty}

\begin{abstract}
Singing voice conversion aims to convert singer's voice from source to target without changing singing content. Parallel training data is typically required for the training of singing voice conversion system, that is however not practical in real-life applications. Recent encoder-decoder structures, such as variational autoencoding Wasserstein generative adversarial network (VAW-GAN), provide an effective way to learn a mapping through non-parallel training data. In this paper, we propose a singing voice conversion framework that is based on VAW-GAN. We train an encoder to disentangle singer identity and singing prosody (F0 contour) from phonetic content. By conditioning on singer identity and F0, the decoder generates output spectral features with unseen target singer identity, and improves F0 rendering. Experimental results show that the proposed framework achieves better performance than the baseline frameworks.


\end{abstract}
\section{Introduction}

Singing voice conversion (SVC) is a voice conversion (VC) technique that converts  source singer's voice to sound like  target singer's voice, while preserving the singing content \cite{villavicencio2010applying}. With singing voice conversion, we can make everyone sing like a professional, overcoming the limitation of physical constraints, controlling the voice timbre freely, and expressing the emotions in variable ways \cite{kobayashi2018intra, luo2020singing}.

Singing voice conversion shares many similarities with speech voice conversion \cite{sisman2020overview, sisman2018voice, fang2018high}. They both aim to change the vocal identity. However, they are also different in many ways. For example, in speech voice conversion, speech prosody is considered to contain the information of speaker characteristics \cite{csicsman2017transformation, sisman2018wavelet, sisman2019group, sisman2017analysis}. In contrast, in singing voice conversion, we assume that source singers are always singing on the key, which means the singing style is only determined by the sheet music, thus we consider singing style as a singer-independent feature. Therefore, only singer-dependent traits, such as vocal timbre, need to be converted \cite{kawakami2010gmm, doi2012, singan-2019}.

Early studies attempted to convert singing voice through spectral modeling.  Many statistical methods, such as Gaussian mixture model (GMM)-based many-to-many eigenvoice conversion (EV-GMM) \cite{toda2007one}, direct waveform modification based on spectrum differential (DIFFSVC) \cite{kobayashi2014statistical}, and DIFFSVC with global variance \cite{kobayashi2015statistical} have been proposed for SVC. With the advent of deep learning, deep neural network (DNN) \cite{hono2019singing} and generative adversarial network (GAN) \cite{singan-2019, sisman2020generative} among others have shown improved quality and naturalness.

\begin{figure}[t]
    \centering
    \includegraphics[scale=0.6]{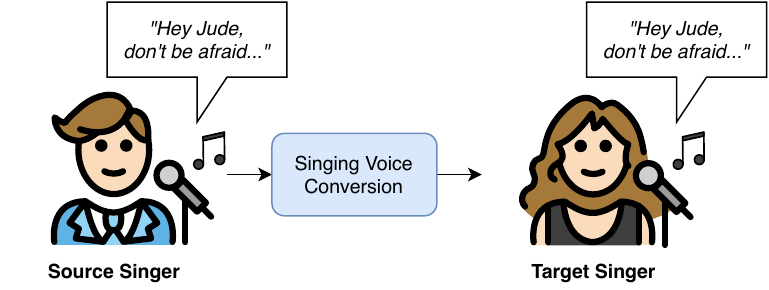}
    \vspace{-3mm}
    \caption{A singing voice conversion system is trained on singing voice data recorded by the source and target singers. At run-time, the system takes singing voice of the source singer as input, and converts it to that of the target singer.}
    \label{fig:intro}
    \vspace{-6mm}
\end{figure}


However, previous studies on singing voice conversion mostly require of parallel training data between the source and target singer. In practice, collecting such parallel data is expensive and time-consuming, that motivated non-parallel SVC methods,
such as deep bidirectional long short term memory (DBLSTM) based recurrent neural network (RNN) \cite{sun2016phonetic, chen2019singing}, Wasserstein generative adversarial network (WD-GAN) \cite{zhao2019singing}, and StarGAN \cite{kameoka2018stargan} for SVC. Recently, encoder-decoder based networks \cite{henter2018deep}, such as variational autoencoder (VAE) \cite{hsu2016voice}, variational autoencoding Wasserstein generative adversarial network (VAW-GAN) \cite{hsu2017}, auxiliary classifier variational autoencoder (ACVAE) \cite{kameoka2018acvae, kameoka2019acvae} and cycle-consistent autoencoder (CycleVAE) \cite{tobing2019non, yook2020many} have been successfully applied into various tasks, such as cross-lingual voice conversion \cite{sismanstudy} and emotional voice conversion \cite{zhou2020converting}.

Autoencoder is effective for disentanglement of mixed information \cite{hsu2016voice, hsu2017,zhou2020seen}. If we are able to disentangle vocal timbre of the singer, that we call singer identity in this paper, from phonetic content and singing prosody (F0 contour), we can simply replace the singer identity and F0 contour during signal reconstruction for singing voice conversion. We adopt VAW-GAN in this paper for two reasons. First, autoencoder doesn't require parallel training data during encoding and decoding; Second, the encoder-decoder architecture allows for effective control of singer identity and singing prosody, that makes many-to-many conversion easier than other non-parallel generative models, such as cycle-consistent generative adversarial network (CycleGAN) \cite{sismanstudy, Zhou2020, fang2018high, lorenzo2018}.

It is known that F0 and spectral features are inherently correlated \cite{kawahara1999restructuring, morise2016world}. In autoencoder-based VC, the latent code from encoder contains F0 information from the source, that adversely affects the output. Recent studies have also shown that disentangling F0 from latent code improves the performance of speech voice conversion \cite{Qian_2020,huang2019} and emotion conversion \cite{zhou2020converting}, that motivates the studies in this paper. We are motivated to study disentanglement of both singer identity, and F0 information from singing content. This can be achieved by providing both singer identity and F0, in addition to latent code, to the decoder during training. In this way, we aim to obtain a latent code that is singer and F0 independent. At run-time, we add both singer identity and F0 as inputs to the decoder to control the signal reconstruction.



The main contributions of this paper include: 1) we propose a framework for singing voice conversion with non-parallel training data; 2) we achieve high quality converted singing voice; 3) we eliminate the need for parallel training data, time alignment procedures and other external modules, such as automatic speech recognition (ASR); 4) we show the effectiveness of the F0 conditioning mechanism for SVC. To our best knowledge, this is the first attempt to use F0 conditioning for non-parallel singing voice conversion.

This paper is organized as follows: In Section II, we recap the study of VAW-GAN for speech synthesis. In Section III, we introduce our proposed singing voice conversion framework. In Section IV, the experiments and results are reported. Conclusions are given in Section V.

\section{Related Work: VAW-GAN in Speech Synthesis}

Recently, encoder-decoder networks such as variational Wasserstein generative adversarial network (VAW-GAN) \cite{hsu2017} have drawn much attention because of their generating ability and controllability.
VAW-GAN makes it possible to train a model without parallel data or any other time alignment procedures through an encoder-decoder structure.

The main idea of VAW-GAN is based on the probabilistic graphical model (PGM). Given spectral features $\bm{x}_{s}$ from  source speaker and $\bm{x}_{t}$ from target speaker, the PGM tries to explain the observation $\bm{x}$ using two latent variables: the speaker representation vector $\bm{y}$ and the phonetic content vector $\bm{z}$. It is noted that $\bm{y}$ is determined solely by the speaker identity and $\bm{z}$ is a speaker-independent variable. According to the PGM, the voice conversion function $f(\cdot)$ can be divided into two stages: 1) a speaker-independent encoder $\mathcal{E}_\phi$ with parameter set $\phi$ infers a latent vector from the source spectral features $\bm{z}=\mathcal{E}_\phi(\bm{x}_{s})$, and 2) a speaker-dependent decoder $\mathcal{G}_{\theta}$ with parameter set $\theta$ reconstructs the input with the latent code $\bm{z}$ and a target speaker representation vector $\bm{y}_t$. Therefore, the task of voice conversion is then reformulated as:
\begin{equation}
    \bm{x}_{t}\approx f(\bm{x}_{s})=\mathcal{G}_{\theta}(\mathcal{E}_\phi(\bm{x}_{s}),\bm{y}_t)
    \label{eq:1}
\end{equation}
During training, the frames that belong to the same phoneme class hinge on a similar $\bm{z}$. With the latent content vector $\bm{z}$, the decoder can generate voice of a specific speaker by varying the speaker representation vector $\bm{y}$.


Different from variational encoding networks, generative adversarial network (GAN) produces sharper spectra since it optimizes a loss function between two distributions in a more direct fashion \cite{hsu2017}. In order to achieve better conversion performance, VAW-GAN incorporates the discriminator from GAN models and assigns VAE’s decoder as GAN’s generator. In the case of voice conversion, the Jensen-Shannon divergence \cite{arjovsky2017wasserstein} in the GAN objective is renovated with a Wasserstein objective:
\begin{equation}
    \mathcal{J}_{wgan}=\mathbb{E}_{\bm{x}\sim p_t^*}[\mathcal{D}_\psi(\bm{x})]-\mathbb{E}_{\bm{z}\sim q_\phi(\bm{z}|\bm{x})}[\mathcal{D}_\psi(\mathcal{G}_\theta(\bm{z}),\bm{y}_t)]
\end{equation}
where $p_t^*$ is the distribution of $\bm{x}_{t}$,
$q_\phi(\bm{z}|\bm{x})$ is the inference model, and $\mathcal{D}_\psi$ is the discriminator with parameter set $\psi$.

Therefore, the final objective loss function of VAW-GAN is given as follow:
\begin{align}
    \mathcal{J}_{vawgan}=
    &-\mathcal{D}_{KL}(q_\phi(\bm{z}|\bm{x})\|p_{\theta}(\bm{z})) \nonumber \\
    &+\mathbb{E}_{\bm{z}\sim q_\phi(\bm{z}|\bm{x})}[\mathrm{log}p_\theta(\bm{x}|\bm{z},\bm{y})]\nonumber \\
    &+\alpha \mathcal{J}_{wgan}
\end{align}
where $\alpha$ is a coefficient which emphasizes $\mathcal{J}_{wgan}$, $\mathcal{D}_{KL}$ is the Kullback-Leibler divergence, $p_\theta(\bm{z})$ is the prior distribution model of $\bm{z}$, and $p_\theta(\bm{x}|\bm{z},\bm{y})$ is the synthesis model. Through the adversarial learning, the decoder minimizes the loss, while the discriminator maximizes it, until an optimal pseudo pair is found through this min-max game. This objective is shared across all three main components in VAW-GAN: the encoder, the decoder and the discriminator.

VAW-GAN has been successfully applied in the field of speech synthesis, such as voice conversion \cite{hsu2017, sismanstudy} and emotion conversion \cite{zhou2020converting}. We expect that the way it characterizes speaker identity also applies to singer identity. In this paper, we propose a VAW-GAN framework for singing voice conversion, which will be the focus of Section III.

\begin{figure*}
    \hspace{10mm}
    \includegraphics[scale=0.75]{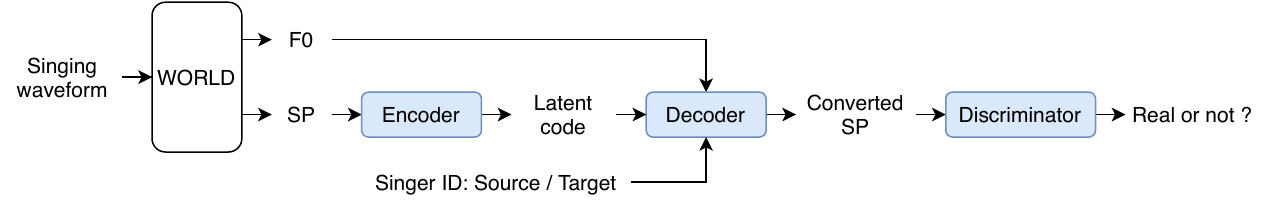}
    \caption{The training phase of the proposed \textit{VAW-GAN (SID+F0)} singing voice conversion framework. The encoder learns to disentangle singer identity and fundamental frequency (F0) from the phonetic content. Blue boxes are involved in the training.}
    \label{fig:training}
\end{figure*}

\begin{figure*}
    \hspace{10mm}
    \includegraphics[scale = 0.75]{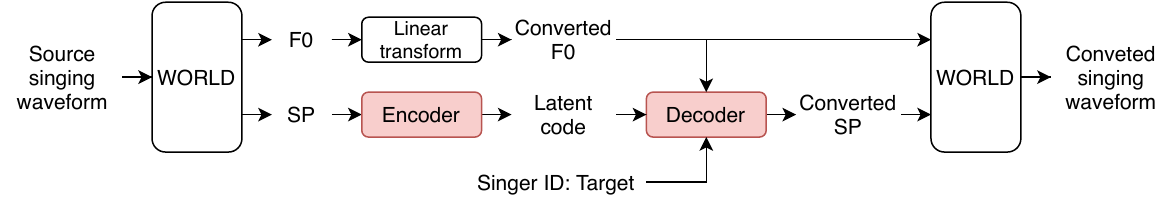}
    \caption{The run-time conversion phase of the proposed \textit{VAW-GAN (SID+F0)} singing voice conversion framework. The decoder is conditioned on singer identity and fundamental frequency (F0) to generate spectral features for unseen target singer, and improve F0 rendering. Red boxes have been trained during the training phase.}
    \label{fig:conversion}
\end{figure*}

\section{VAW-GAN for Singing Voice Conversion}
In this section, we propose the use of VAW-GAN for disentanglement of singer identity and F0 information from the phonetic content. The proposed VAW-GAN includes a singer-independent encoder, that generates latent code $\bm{z}$, and a decoder that takes a triplet input, namely latent code, singer identity and F0.


\subsection{Training Phase}
The training phase is illustrated in Fig. \ref{fig:training}. We first use WORLD vocoder \cite{morise2016world} to extract spectral features (SP) and F0 from the singing waveform. The encoder takes input frames from multiple singers, and generate a singer-independent latent code $\bm{z}$. We assume that the latent code $\bm{z}$ only contains the information of phonetic content.

We use a one-hot vector singer ID and source F0 as the input to the decoder, in addition to latent code. In this way, the encoder learns to disentangle singer ID and F0 from the latent code $\bm{z}$ after being exposed to singing data of multiple singers. By conditioning on singer ID and F0, the decoder, as formulated in Eq. (\ref{eq:1}), can be re-written as follows,
\begin{equation}
    \bm{x}\approx f(\bm{x},F0)=\mathcal{G}_{\theta}(\mathcal{E}_\phi(\bm{x}),\bm{y},F0)
\end{equation}

The decoder learns to reconstruct the spectral features and the discriminator tries to distinguish whether the spectral features are from real singing voice or not. Through this min-max game, the encoder, decoder and discriminator are encouraged to find an optimal pseudo pairs during the training.




\subsection{Run-time Conversion}

The conversion phase is illustrated in Fig. \ref{fig:conversion}. We first extract spectral features and F0 from the source singing waveform using WORLD vocoder. The spectral features are then encoded into a latent code through the encoder and F0 is converted by logarithm Gaussian (LG)-based linear transformation \cite{hsu2016voice}.

At run-time  conversion, the decoder is conditioned on the converted F0 features $\hat{F0}$. The converted spectral features $\hat{\bm{x}}_{t}$ are given as:
\begin{equation}
    \hat{\bm{x}}_{t}=f(\bm{x}_{s},\hat{F0})=\mathcal{G}_{\theta}(\mathcal{E}_\phi(\bm{x}_{s}),\bm{y}_t,\hat{F0})
\end{equation}
where $\bm{y}_t$ is the designated singer ID.

The converted spectral features are then reconstructed by the decoder together with the converted F0 and the designated singer ID. Finally, we use WORLD vocoder to synthesis the converted singing waveform.


\begin{table*}[ht]
    \caption{The model architecture of the encoder, decoder and discriminator of our proposed framework \textit{VAW-GAN (SID+F0)}.}
    \centering
    \begin{tabular}{|c|c|c|c|c|}
    \hline
                    &  \textbf{ \# of Layers}    &  \textbf{Kernel Size}      &  \textbf{Stride}           &  \textbf{Output Channel}           \\ \hline
    Encoder         &  5        &  \{7, 7, 7, 7, 7\}  &  \{3, 3, 3, 3, 3\}  & \{16, 32, 64, 128, 256\}   \\ \hline
    Decoder         &  4        &  \{9, 7, 7, 1025\}  &  \{3, 3, 3, 1\}    &  \{32, 16, 8, 1\}         \\ \hline
    Discriminator   &  3        &  \{7, 7, 115\}      &  \{3, 3, 3\}       &  \{16, 32, 64\}            \\ \hline
    \end{tabular}
    \label{tab:architecture}
\end{table*}

\section{Experiments}

We conduct both objective and subjective experiments to assess the performance of the proposed VAW-GAN for singing voice conversion. We use NUS Sung and Spoken Lyrics Corpus (NUS-48E corpus) \cite{singing_data}, which consists of the sung and spoken lyrics of 48 English songs by 12 professional singers. We choose 2 male singers and 1 female singer for all the experiments.
For each singer, 6 songs are used for training and evaluation.

We construct two systems: a) VAW-GAN with the decoder conditioning on singer ID (SID) and F0 (as illustrated in Figure \ref{fig:conversion}) to convert the spectrum between different singers, denoted as \textit{VAW-GAN (SID+F0)}; b) VAW-GAN with the decoder conditioning only on singer ID, denoted as \textit{VAW-GAN (SID)}, that is similar to the VAW-GAN in~\cite{hsu2017} for speech voice conversion.

We use \textit{VAW-GAN (SID)} as the reference baseline to show the effect of the proposed~\textit{VAW-GAN (SID+F0)}, and report the performance in both objective and subjective evaluations. It is noted that both frameworks are trained with non-parallel singing voice data.


\subsection{Experimental Setup}


The singing voice data is down-sampled at 16kHz. We first use WORLD vocoder \cite{morise2016world} to extract 513-dimensional spectral features (SP), F0, and aperiodicity (AP). The frame length is 25 ms  with a frame shift of 5 ms. F0 is re-scaled to the range of $[-1,1]$. The input SP of each frame is normalized to unit-sum. The normalization factor, known as the energy, is taken out as an independent feature. We use the log energy-normalized SP for VAW-GAN training.


\begin{table}[t]
\caption{A comparison of the MCD results between VAW-GAN (SID+F0), VAW-GAN (SID) for male-to-male and male-to-female singing voice conversion. }
\centering
\begin{tabular}{c||c|c}
\hline
\multirow{2}{*}{Framework} & \multicolumn{2}{c}{MCD {[}dB{]}} \\ \cline{2-3}
                           & male $\rightarrow$ male      & male $\rightarrow$ female      \\ \hline
Zero effort                & 10.05          & 13.43            \\ 
VAW-GAN (SID) & 7.20           & 7.39             \\ 
\textbf{VAW-GAN (SID+F0)}      & \textbf{5.51}           & \textbf{6.57}             \\ \hline
\end{tabular}
\label{tab:mcd}
\end{table}

The model architecture of our proposed \textit{VAW-GAN (SID+F0)} framework is given in Table \ref{tab:architecture}. The encoder, the decoder, and the discriminator of both frameworks are all 1D convolutional neural networks (CNN), of which each layer is followed by a fully connected layer. The latent space is 128-dimensional and is assumed to have a standard normal distribution. The dimension of the speaker representation is set to be 10.  
For both frameworks, during the training phase, we first train the encoder-decoder pair for 15 epochs, and then train the whole VAW-GAN for 45 epochs. The framework is trained using RMSProp with a learning rate of 0.0001. During the conversion phase, SP and F0 are converted on a frame-by-frame basis, while AP and the energy remain unmodified.



\subsection{Objective Evaluation}

We use Mel-cepstral distortion (MCD) \cite{mcd1,kobayashi2018intra} to measure the distortion between the converted and target Mel-cepstra, that is given as follows:

\begin{equation}
    MCD[dB]=\frac{10}{T\ln{10}}\sum_{i=1}^{T}\sqrt{2\sum_{d=1}^{D}(mc_{i,d}^{t}-mc_{i,d}^{c})^2}
\end{equation}
where $mc_{i,d}^{c}$ and $mc_{i,d}^{t}$ represent the $d^{th}$ coefficient of the converted and target MCEPs sequences at the $i^{th}$ frame, respectively. $D$ is the dimension of MCEP features and $T$ represents the total number of frames. In this paper, we extract 24-dimensional MCEPs at each frame, thus $D$ is 24. We note that a lower value of MCD indicates a smaller spectrum distortion and a better conversion performance.

We report MCD results of our proposed framework \textit{VAW-GAN (SID+F0)} and the baseline framework \textit{VAW-GAN (SID)}. Zero effort represents the cases where we directly compare the singing voice of source and target singers without any conversion. As reported in Table \ref{tab:mcd}, we observe that our proposed framework \textit{VAW-GAN (SID+F0)} outperforms the baseline framework \textit{VAW-GAN (SID)} for both male-to-male and male-to-female (5.51 vs. 6.20 and 6.57 vs. 7.39), which we believe is remarkable. The results indicate that our proposed \textit{VAW-GAN (SID+F0)} with F0 conditioning achieves a better performance of spectrum conversion than the baseline framework \textit{VAW-GAN (SID)} without any condition in both inter-gender and intra-gender SVC.





\begin{figure}[h]
    \includegraphics[scale=0.55]{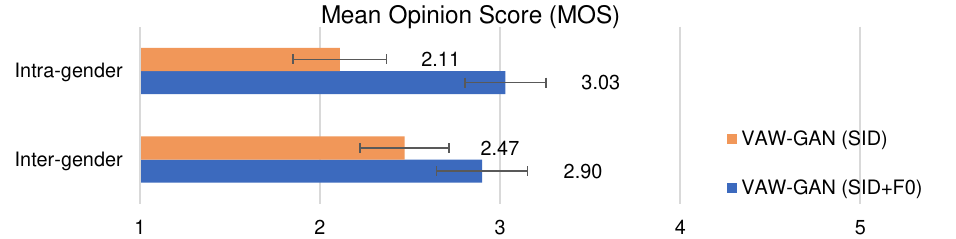}
    \caption{MOS results with 95 \% confidence interval between the proposed \textit{VAW-GAN (SID+F0)} and \textit{VAW-GAN (SID)} baseline.} 
    \label{fig:mos}
\end{figure}

\begin{figure}[h]
    \includegraphics[scale=0.55]{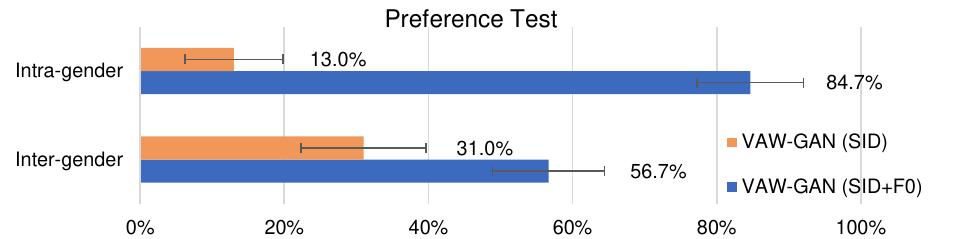}
    \caption{XAB preference test results with 95 \% confidence interval between the proposed \textit{VAW-GAN (SID+F0)} and \textit{VAW-GAN (SID)} baseline.}
    \label{fig:preference}
\end{figure}

\subsection{Subjective Evaluation}

We further conduct subjective evaluation to assess the performance of the proposed \textit{VAW-GAN} for singing voice conversion in terms of voice quality and singer similarity. 20 subjects participate in all the listening tests, and each of them listens to 120 converted singing voice samples in total.

We conduct mean opinion score (MOS) \cite{streijl2016mean, liu2019teacher} to assess the voice quality of the converted singing voices.
Listeners are asked to score the quality of the converted singing voice on a five-point scale (5: excellent, 4: good, 3: fair, 2: poor, 1: bad).
As shown in Fig. \ref{fig:mos}, our proposed framework \textit{VAW-GAN (SID+F0)} outperforms the baseline framework \textit{VAW-GAN (SID)} in terms of voice quality by achieving higher MOS values of 3.03 $\pm$ 0.28 for male-to-male singing voice conversion and 2.90 $\pm$ 0.31 for male-to-female singing voice conversion. The results suggest that conditioning the decoder on F0 improves voice quality remarkably, which is consistent with the observation in objective evaluation.

We also conduct XAB preference test \cite{helander2010voice, liu2020wavetts} in terms of the singer similarity. The subjects are asked to listen  to  the  reference target singing samples and the converted singing samples of the \textit{VAW-GAN (SID)} baseline, and the proposed \textit{VAW-GAN (SID+F0)}, and  choose  the  one  which  sounds closer to the target in terms of singer similarity.
As shown in Fig. \ref{fig:preference}, our proposed framework \textit{VAW-GAN (SID+F0)} outperforms the \textit{VAW-GAN (SID)} baseline in terms of singer similarity (84.7 \% vs. 13 \% for male-to-male SVC and 56.7 \% vs. 31 \% for male-to-female SVC). The results prove the effectiveness of our proposed framework in terms of singer identity conversion.

\section{Conclusion}

In this paper, we propose a parallel-data-free singing voice conversion framework with VAW-GAN. We first propose to conduct singer-independent training with a encoding-decoding process. We also propose to condition the decoder on F0 to improve the singing voice conversion performance. We eliminate the need for parallel training data or other time-alignment procedures, and achieve high performance of converted singing voices. Experimental results show the efficiency of our proposed SVC framework on both intra-gender and inter-gender singing voice conversion.

\section{Acknowledgement}

This work is supported by the National Research Foundation, Singapore under its AI Singapore Programme (AISG Award No: AISG-100E-2018-006, AISG-GC-2019-002),  the National Robotics Programme (Programmatic Grant Number: 192 25 00054), Human Robot Collaborative AI for AME Programmatic Grant (Programmatic Grant No. A18A2b0046) and Neuromorphic Computing Programmatic Grant (Programmatic Grant No. A1687b0033) from the Singapore Government’s Research, Innovation and Enterprise 2020 plan in the Advanced Manufacturing and Engineering domain. Any opinions, findings and conclusions or recommendations expressed in this material are those of the author(s) and do not reflect the views of National Research Foundation, Singapore.

This work is also supported by SUTD Start-up Grant Artificial Intelligence for Human Voice  Conversion  (SRG  ISTD  2020  158)  and  SUTD  AI  Grant  titled 'The Understanding and Synthesis of Expressive Speech by AI' (PIE-SGP-AI-2020-02).

\bibliographystyle{IEEEtran}
{\footnotesize
}
\end{document}